\documentclass{amsart}
\usepackage{amssymb}
\usepackage{amsfonts}
\usepackage{amsmath}
\usepackage[legalpaper,bookmarks=true,colorlinks=true,linkcolor=blue,citecolor=blue]{hyperref}
\usepackage{graphicx}%
\usepackage{fancyhdr}
\usepackage{color}
\usepackage[mathlines]{lineno}
\usepackage{lscape}
\usepackage{epsfig}
\usepackage{natbib}
\usepackage{geometry}
\usepackage{tgbonum}
\usepackage[]{listings}
\fontfamily{qcr}\selectfont
\newtheorem{theorem}{Theorem}
\theoremstyle{plain}

\newtheorem{corollary}{Corollary}

\numberwithin{equation}{section}

\newcommand{\Bin}{\bigskip \noindent}

\newcommand{\Ni}{\noindent}
\newcommand{\vb}{|}

\begin{document}
\Large
\title{A Jarque-Bera test for skew normal data}
\author{Diam Ba $^{(1)}$}
\author{Gorgui Gning $^{(2)}$}
\author{Gandasor Bonyiri Onesiphore Da $^{(3)}$}
\author{Oumar Foly Sow $^{(4)}$}
\author{Gane Samb Lo $^{(5)}$}
\maketitle

\noindent $^{(5)}$ Gane Samb Lo.\\
LERSTAD, Gaston Berger University, Saint-Louis, S\'en\'egal (main affiliation).\newline
LSTA, Pierre and Marie Curie University, Paris VI, France.\newline
AUST - African University of Sciences and Technology, Abuja, Nigeria\\
Imhotep International Mathematical Center (imh-imc), https://imhotepsciences.org\\
gane-samb.lo@edu.ugb.sn, gslo@aust.edu.ng, ganesamblo@ganesamblo.net\\
Permanent address : 1178 Evanston Dr NW T3P 0J9, Calgary, Alberta, Canada.\\

\noindent $^{(1)}$ Dr. Diam Ba\\
Ecole Normale Supérieure d'Enseignement Technique\\
professionnel, ENSETP\\
University Cheikh Anta Diop, Dakar, SENEGAL\\
Imhotep International Mathematical Center (imh-imc), https://imhotepsciences.org\\
Research affiliation : LERSTAD, Gaston Berger University, Saint-Louis, SENEGAL\\
Emails : diam.ba@ucad.edu.sn, diamba79@gmail.com\\

\noindent $^{(2)}$ Dr. Gorgui Gning\\
University of Labé, Sciences  Faculty, GUINEA\\
Imhotep International Mathematical Center (imh-imc), https://imhotepsciences.org\\
Researcher at LERSTAD, Gaston Berger University, Saint-Louis, SENEGAL.\\
Email: \\

\noindent $^{(3)}$ Gandasor Bonyiri Onesiphore Da\\
Imhotep International Mathematical Center (imh-imc), https://imhotepsciences.org\\
Researcher at LERSTAD, Gaston Berger University, Saint-Louis, SENEGAL.\\
Email: da.gandasor-bonyiri-onesiphore@ugb.edu.sn, gandasor@gmail.com\\

\noindent $^{(4)}$ Oumar Foly Sow (M.Sc)\\
University of Labé, Sciences Faculty, GUINEA\\
oumar-foly.sow@univ-labe.edu.gn\\
oumarfolysow363@gmail.com\\

%\noindent Dr. Aladji Babacar Niang\\
%University of N'zérékoré, Sciences and Techniques Faculty, GUINEA\\
%Imhotep International Mathematical Center (imh-imc), https://imhotepsciences.org\\
%Researcher at LERSTAD, Gaston Berger University, Saint-Louis, SENEGAL.\\
%Email: niang.aladji-babacar@ugb.edu.sn, aladjibacar93@gmail.com\\

%\noindent Abdoulaye Condé (M.Sc)\\
%University of N'zérékoré, Sciences Faculty, GUINEA
%abdoulayeconde2017@gmail.com\\
%Email: oumarfolysow363@gmail.com\\

\newpage
\Ni \textbf{Full Abstract}  \label{2024_03_01_abs} The skew normal law has been introduced in Azzalin (1985) as an alternative to adjusting asymmetric data that share important patterns with the normal law. It has been extensively studied. However, there is so much to do in order to catch the diversity and the richness of the investigation of its normal counterpart. The General Jarque-Berra Test (GJBT) has been devised by Lo et al. (2015), Da et al. (2023) for arbitrary laws with at least finite first eight moments, as a generalization of the Jarque-Bera (1987) test that was specially set up for normal data. Here, we particularize it to skew normal data.  When particularized in the skew normal law, this test is proven to be extremely powerful in detecting the true model for any $\alpha\neq 0$ and rejected the normal law ($\alpha=0$) whatever be the size of the data. We introduced the use of the samples duplication method to reach a high level of efficiency for the test.\\

\Ni \textbf{Résumé} (Full Abstract in French) La loi normale asymétrique (skew normal law) a été introduite dans Azzalin (1985) comme une alternative pour modéliser des données asymétriques partageant des caractéristiques importantes avec la loi normale. Elle a fait l'objet de nombreuses études approfondies. Toutefois, beaucoup reste à faire pour atteindre la diversité et la richesse des investigations menées sur la loi normale. Le test généralisé de Jarque-Bera (GJBT) a été proposé par Lo \textit{et al.} (2015), puis par Da \textit{et al.} (2023), pour des lois arbitraires possédant au moins les huit premiers moments finis, en tant que généralisation du test de Jarque-Bera (1987) spécifiquement conçu pour les données normales. Dans ce travail, nous nous intéressons au cas particulier des données suivant une loi normale asymétrique. Il est démontré que ce test présente une puissance remarquable pour détecter le véritable modèle pour tout $\alpha \neq 0$, en rejetant systématiquement la loi normale ($\alpha = 0$), quelle que soit la taille de l'échantillon. Nous introduisons également l'utilisation de la méthode de duplication d'échantillons afin d'accroître significativement l’efficacité du test.\\

\newpage
\Ni \textbf{Presentation of authors}.\\

\Ni \textbf{Diam Ba}, Ph.D., is a Senior Lecturer of Statistics and Mathematics at the University of Cheikh Anta Diop, Senegal. He is a lead researcher at: Imhotep International Mathematical Center (imh-imc), https://imhotepsciences.org.\\

\Ni \textbf{Gorgui Gning}, Ph.D., is a Lecturer at the University of Lab\'e, Lab\'e, Guin\'ee. He is a lead researcher at: Imhotep International Mathematical Center (imh-imc), https://imhotepsciences.org.\\

\Ni \textbf{Gandasor Bonyiri Onesiphore Da}, M.Sc, is preparing a Ph.D. degree at: Doctoral Studies, LERSTAD, Universit\'e Gaston Berger, Saint-Louis, SENEGAL, and in the Imhotep International Mathematical Center (imh-imc), https://imhotepsciences.org, under the supervision of Gane Samb Lo and Tchilabalo Abozou Kpanzou.\\

\Ni {\bf Oumar Foly Sow}, M.Sc., is a lecturer at University of Labé, Sciences Faculty, GUINEA\\

\Ni \textbf{Gane Samb Lo}, Ph.D. and Doctorat ès Sciences, is a retired full professor from Université Gaston Berger (2023), Saint-Louis, SENEGAL. He is the founder and The Probability and Statistics Chair holder at: Imhotep International Mathematical Center (imho-imc), https://imhotepsciences.org.\\

\section{Introduction}

\noindent The skew normal statistical distributions form class of probability laws on $\mathbb{R}$ that include the normal laws. It has been  introduced by \cite{azzalini1}. Since then, that class attracted many researchers who tried to extend the properties of Gaussian law  to non-symmetric laws but presenting light tail. Skew normal laws are used to model real life data in a quite large number of fields.  To cite a few, \cite{kim} and \cite{counsell} applied them to psychiatric data, \cite{carm} to finance prices, etc. Also, like for the normal law that has been used to generate variance mixing laws, a number of mixtures of the skew normal has been set up and applied to finance data. In summary, this model has been studied, both in probability theory and in Statistical modeling in an important number of directions.  \\

\noindent Yet, there are many things to be studied about it and refinements are highly expected. Here we focus on goodness-of-fit statistical tests for skew normal distribution. By making profit of the generalized Jarque-Bera test of \cite{gsloJB2015} refined by \cite{da}, we deal, here, on with a particular form related to the skew normal law. By doing so, we extend the Jarque-Bera test for the symmetric normal law to another Jarque-Bera test for the asymmetric skew normal law.

\noindent Now, to put the reader in his ease, we wish to provide a pedagogical introduction to our studied law.

\section{A quick round up of the standard Skew Normal law} \label{sec2}

 The normal skewness is a continuous probability distribution function which generalizes the normal law by introducing a non null skewness. Given the probability distribution function (\textit{pdf}) and the cumulative distribution function (\textit{cdf}) of the standard  normal law defined as
 
$$
\phi(x)=\frac{1}{\sqrt{2\pi}}e^{-\frac{x^2}{2}},\ x\in\mathbb{R}, \ \ \Phi(x)=\int_{-\infty}^{x}\phi(t)dt.
$$
	
\Bin and given a real-value parameter $\alpha$, the following function

\begin{equation}\label{skew_01}
f_{\alpha}(x)=2\phi(x)\Phi(\alpha x), \ \ x \in \mathbb{R}.
\end{equation}

\Bin is a \textit{pdf} whose corresponding probability law is called as the standard \textit{skew normal law}, denoted as 
$SN(\alpha,1,0)$. The general skew normal random variable is defined as $SN(\alpha,a,b)=a SN(\alpha,1,0)+b$, with $a>0$, $b\in \mathbb{R}$. For short, $SN(\alpha,1,0)$ is simply denoted as $SN(\alpha)$.\\

\begin{figure}
\centering
			\includegraphics[width=1.00\textwidth]{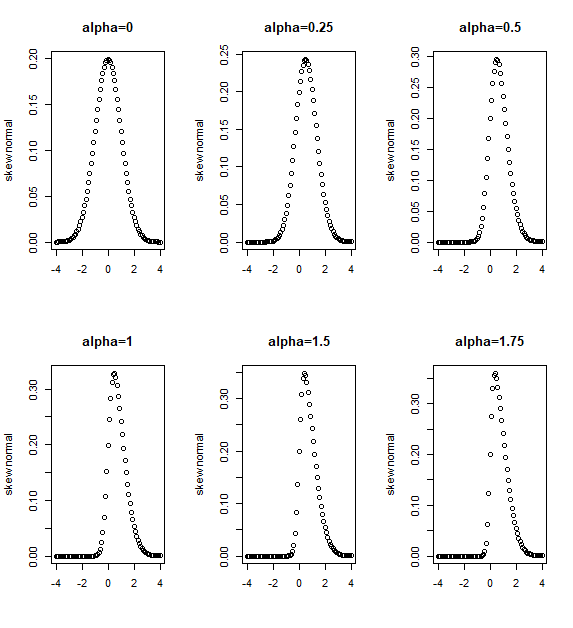}
			\caption{Displacement to left of the mean for $\alpha>0$}
	\label{figSN1}
\end{figure}

\begin{figure}
	\centering
		\includegraphics[width=1.00\textwidth]{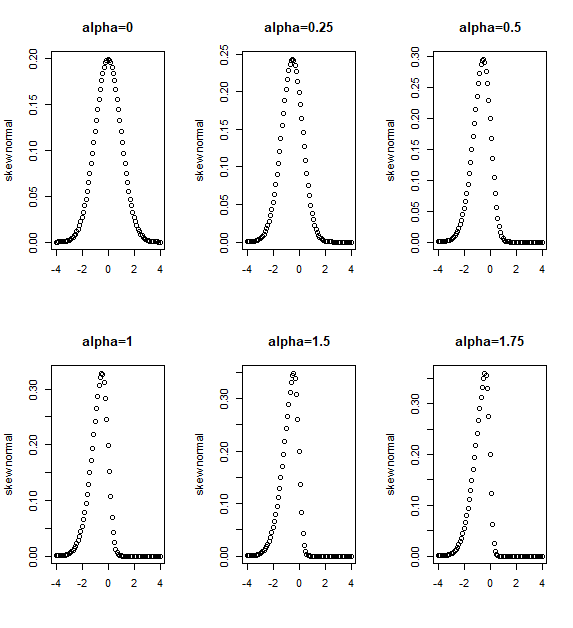}
		\caption{Displacement to right of the mean for $\alpha<0$}
		\label{figSN2}
\end{figure}

\Ni The pivotal property of skew normal laws is that a non-null value $\alpha$ makes the distribution non-symmetrical and a null-value make the distribution to be a classical normal law. It is crucial to notice that the transform focuses on the introduced asymmetry and not on the shape of \textit{pdf}. We can see the drift from the asymmetry of $SN(\alpha,1,0)$ led by  $\alpha>0$ in Fig. \ref{figSN1} and 
by $\alpha<0$ in Fig. \ref{figSN2}. For positive values, the center of the original normal law is displaced to left and to right for negative values.

\Bin Now we are going to give remarkable properties of that law. One of the most interesting one, if not the most important, is the data generation method given in Fact 3 below.\\

\noindent \textbf{Facts 1. Boundary laws}.\\

\Ni (1) $SN(0,1,0)=\mathcal{N}(0,1)$; As $\alpha\rightarrow 0$, then $SN(\alpha,1,0)  \rightsquigarrow  \mathcal{N}(0,1)$.\\

\Ni (2) When $\alpha\rightarrow +\infty$ (resp. $\alpha\rightarrow -\infty$), then $SN(\alpha,1,0)  \rightsquigarrow  |\mathcal{N}(0,1)|$ (resp. $SN(\alpha,1,0)  \rightsquigarrow -|\mathcal{N}(0,1)|$).\\

\Ni \textbf{Facts 2. Law transformation}. We have\\

\Ni (3) For any $\alpha \in \mathbb{R}$, $-SN(\alpha)=_d-SN(-\alpha)$.\\

\Ni (4) $SN(\alpha)$  and $\vb SN(\alpha)\vb$ are independent.\\

\Ni (5) $SN(\alpha)^2 \sim \chi^2_1$.\\

\Ni (6) If $Z \sim \mathcal{N}(0,1)$ and independent of $X\sim SN(\alpha)$, then

$$
\frac{X+Z}{2} \sim SN\biggr( \frac{\alpha}{\sqrt{2+\alpha^2}}\biggr).
$$

\Bin  \textbf{Consequence}. Because of Point (3) above, we usually restrict the study of skew normal laws to $\alpha>0$. $\square$\\

\newpage
\Ni \textbf{Facts 3. Generation}. Let $\alpha \in \mathbb{R}$ and $\delta=\alpha(1+\alpha^2)^{-1/2}$. Let $Z_i \sim \mathcal{N}(0,1)$, $i-1,2$, be two independent standards normal laws. We can generate $SN(\alpha)$ as 

$$
\delta |Z_1| +(1-\delta^2)^{1/2} Z_2 \ \sim \ SN(\alpha,0,1).
$$

\Ni \textbf{Proofs}. They are to be found in some papers, in \cite{mohsen} and \cite{haas} in particular.\\

\Ni The moments of skew normal laws are also largely studied in the literature. Since these are strongly involved in the topic of Generalized Jarque-Bera (GJB) tests, we postpone their exposure in the next section.\\

\Ni Also since the Generalized Jarque-Bera (GJB) test is either quite new or not well spread yet, we are giving it in the new version or \cite{da} at order $p=2$.\\

\section{Recalls on Generalized Jarque-Bera (GJB) test}

\Ni We are going to adapt the \textit{GJB} test as refined in \cite{daGGMG2023}. We begin by describing the method with recalls. For a reminder, the classical Jarque-Bera test belongs to the class of omnibus tests, i.e those which assess simultaneously whether the skewness and kurtosis are consistent with a Gaussian model. This test proved optimum asymptotic power and good finite sample properties (see \cite{jarque}).\\

\Ni A detailed description of that test and related indepth analyses can be found in \cite{bowman}, \cite{dagostino} etc. 
So we begin by describing that classical Jarque Bera test.\\

\Ni Let $X_{1}$, $X_{2}$, ... be a sequence of independent real random variables defined on the probability space  with the same distribution $F$. We suppose that $F$ has a finite eight-th moment 

\begin{equation*}
m_{8}=\int_{\mathbb{R}}\left|x\right|^{8}dF(x)< \infty.
\end{equation*}

\Bin Let us consider the empirical and exact kurtosis and skewness parameters of $F$

\begin{equation*}
a_{n}=\frac{1/n \sum_{i=1}^{n}(X_{i}-\bar{X})^{4}}{1/n \sum_{i=1}^{n}((X_{i}-\bar{X})^{2})^2},
\end{equation*}

\Bin

\begin{equation*}
b_{n}=\frac{1/n \sum_{i=1}^{n}(X_{i}-\bar{X})^{3}}{\left[1/n \sum_{i=1}^{n}(X_{i}-\bar{X})^{2}\right]^{3/2}},\\
 \end{equation*}

\Bin

\begin{equation*}
a=\frac{\mathbb{E}(X-\mathbb{E}X)^4}{\sigma^4}  \ \ and \ \ b=\frac{\mathbb{E}(X-\mathbb{E}X)^3}{\sigma^3}
\end{equation*}

\Bin with $\sigma^2=\mathbb{E}(X-\mathbb{E}X)^2$ is the variance of $X$. The statistic of Jarque-Bera for testing the normal distribution for which $a=3$ and $b=0$, is the following 

\begin{equation*}
J_{n}=\frac{n}{6}\left(\frac{1}{4}(a_n-3)^{2}+b_{n}^{2}\right).
\end{equation*}

\Bin The test is one of the most powerful tools for testing normality of a sample. Empirical studies show that this test accepts normality and rejects non normality for sizes around $n=22$  and $n=25$.  It is obvious that this Jarque-Bera test is based on the first fourth moments 

\begin{equation*}
m_{j}=\int_{\mathbb{R}}\left|x\right|^{j}dF(x), \ \ j=1,...,4.
\end{equation*}

\Bin Since it is derived from the asymptotic normality status of $a_n$ and $b_n$ the moment $m_j$, $j=5,...,8$ are also required. This secret as revealed by \cite{gsloJB2015} showed that the Jarque-Bera test really relies  on the first eight moments and this might explain how much it is efficient for detecting normality and rejecting non normality.\\

\Ni The theorems below justify the Generalize Jarque-Bera test based on the eight first moments:

$$
m_0=1, \ \ m=:m_1=\mathbb{E}X \ \ and \ \ m_j=\mathbb{E}X^{j}, \ 1\leq 2\leq 8. 
$$

\Bin and the functions $h_0(x)=1$, $h_j(x)=x^j$, $j\geq 1$.\\

\Ni We first havethe  asymptotic normality law of the empirical skewness and kurtosis.

\begin{theorem}\label{theo1}
\Ni Let us suppose that the eight-th moments exist. Then we have, as $n \rightarrow +\infty$,

\begin{eqnarray*}
\sqrt{n}\left(( a_n-a), \  ( b_n-b)\right)&=&(\mathbb{G}_{n}(C), \ \mathbb{G}_{n}(B))+o_{\mathbb{P}}(1)\\
&\rightarrow& Z=(Z_1, \ Z_2) \sim\mathcal{N}((0,0),\Sigma)
\end{eqnarray*}

\Bin where $\Sigma_{1,1}=\mathbb{V}ar(C(X))$, $\Sigma_{2,2}=\mathbb{V}ar(B(X))$ and  $\Sigma_{1,2}=\Sigma_{2,1}=Cov(C(X), B(X))$ and

\begin{eqnarray} \label{A2}
&&\textcolor{white}{x} \notag\\
&&(m_2-m_1^2)^4 C\\
&&=(m_2-m_1^2)^2 h_4-4m_1 (m_2-m_1^2)^2 h_3 \notag \\
&&+\left\{6m_1^2(m_2-m_1^2)^2-2(m_2-m_1^2)(m_4-4m_1m_3+6m_1^2m_2-3m_1^4)  \right\}h_2 \notag \\
&&+\left\{(m_2-m_1^2)^2(-4m_3+12m_1m_2-12m_1^3) \notag \right.\\
&&\left. +4m_1(m_2-m_1^2)(m_4-4m_1m_3+6m_1^2m_2-3m_1^4)\right\}h_1 \notag 
\end{eqnarray}

\Bin and

\begin{eqnarray} \label{B2}
&&\textcolor{white}{x} \notag\\
&&(m_2-m_1^2)^3 B\\
&&= (m_2-m_1^2)^{3/2}h_3 +\left\{-3m_1(m_2-m_1^2)^{3/2} \right. \notag \\
&& \left. -(3/2)(m_3-3m_1m_2+2m_1^3)(m_2-m_1^2)^{1/2}\right\}h_2 \notag\\
&&+\left\{(m_2-m_1^2)^{3/2}(-3m_2 + 6m_1^2) \right. \notag\\
&& \left. +3m_1(m_3-3m_1m_2+2m_1^3)(m_2-m_1^2)^{1/2}\right\}h_1. \notag
\end{eqnarray}
\end{theorem}

\Bin From this theorem , we can derive the following corollary:

\begin{corollary}\label{col1} Under the hypothesis of the theorem, we have the general Jarque-Bera test:\\

\Ni (1) If $X$ is symmetrical, then $Z_1$ and $Z_2$ are independent with $det(\Sigma)=0$ and

\begin{equation*}
J_{n}=n \ \Sigma_{11}^{-1}(a_{n}-a)^{2}+\Sigma_{22}(b_{n}-b)^{2} \rightarrow \chi_{2}^{2},
\end{equation*}

\Bin (2) If $X$ is not symmetrical, and if $det(\Sigma)\neq0$, i.e., $Z_1$ and $Z_2$ are not independent, then

\begin{equation*}
J_{n}=n \ \left[\Sigma_{22}(a_{n}-a)^{2}+\Sigma_{11}(b_{n}-b)^{2}-2\Sigma_{12}(a_n-a)(b_n-b) \right]/\det(\Sigma) \rightarrow \chi_{2}^{2},
\end{equation*}
\end{corollary}

\Bin Let us give a  simple example:\\

\Ni  \textbf{Classical Jarque-Bera test with a symmetrical law}. Suppose that the data is symmetrical or at least null first odd moments: 

$$
m_{2j+1}=0, \ j\in {0,1,2,3},
$$

\Bin Then we get

$$
C=-6h_{2}+h_{4}    \ \  and \ \ B=-3h_{1}+h_{3}.
$$  

\Bin \Ni  \textbf{Classical Jarque-Bera test with a Gaussian law}. Now, when particularized for normal laws, we get

$$
\sigma_{11}= 24, \ \ \sigma_{22}=6 \ \ and \ \sigma_{12}= \sigma_{21}=0.
$$

\Bin We see that we get again the asymptotic law of $J_n$ in the normal case, that is the classical Jarque-Bera test.\\

\Ni \textbf{An important remark}. It is clear that the \textit{GJBT} and all the constants in the results are invariant in the normalization and localization coefficients. So adjusting $Y=AX+B$ and  adjusting $X$ are exactly the same. So, we develop the rest of the paper for 
$X \sim SN(\alpha, 0,1)$.\\

\section{Results of the application of the GJBT to  skew normal laws}

\subsection{On the moments of Skew Normal laws} 

\Ni  We begin to handle the pure model $X \sim SN(\alpha, 0,1)$. We need the first eight non-centered moments. General formulas of the centered moments are given in \cite{martinez} and we recall them, here, for the readers: 

$$
\forall n \geq 1, \ \mu_{2n}=\mu_{2n}\biggr(\mathcal{N}(0,1)\biggr), 
$$ 

\Bin i.e., the standard skew and symmetrical normal laws have the same even moments, and
$$
\mu_{2n+1}=\sqrt{\frac{2}{\pi}}\frac{\alpha}{(1+\alpha^{2})^{(2n+1)/2}}\sum_{k=0}^{n}\sum_{j=0}^{n-k}\frac{(2n)!!}{(2k)!!}(2k-1)!!\alpha^{2j}.
$$

\bigskip \noindent where 
 
$$
n!!=\prod_{j \geq 0}(n-2j), \ \ n-2j>0.
$$

\Bin However we do not rely on these formulas. We use the representation of $SN(\alpha,0,1)$ law as $Z=A Z_1 + BZ_2$, where 
$Z_1$ and $Z_2$ are independent, $Z_1 \sim |\mathcal{N}(0,1)|$ and $Z_2 \sim \mathcal{N}(0,1)$,  $A=\alpha(1+\alpha^2)^{-1/2}=\delta$ and $B=\sqrt{1-A^2}$ so that, $A \in [0,1[$, $A^2+B^2=1$. The first eight moments moments $M1[k] = \mathbb{E}(Z_1^k)$, for  $k \in \{0,\cdots,8\}$ can be easily, as shown in Appendix (M1) in page \pageref{anmoments}, as

$$
M_1=(\mathbb{E}Z_1^j, \ 0\leq j \leq 8)= (1, c, \ \ 1, \ \  2c, \ \  3, \ \  8c, \ \  15, \ \  48c ,\ \ 105)
$$

\Bin The corresponding moments for $Z_2$ are

$$
M_2=(\mathbb{E}Z_2^j, \ 1\leq j \leq 8)= (0, \ \ 1, \ \  0, \ \  3, \ \  0, \ \  15, \ \  0 ,\ \ 105).
$$

\Bin Then for the purpose of the implementation of the \textit{GJB} test, we can use binomal expansions to performs statistical tests.  Nevertheless, closed-forms expression in $A$ and $B$ - where the fact $A^2+B^2=1$ is used - or in $\alpha$ are still used for theoretical interpretations. Direct computations using combinatorial methods and the values of matrices $M_1$ and $M_2$ lead to: 
\newpage
\begin{eqnarray*}
c&=&\sqrt{2/\pi}\\
&&\\
M[1]&=&Ac\\
&=& c \left\{ \frac{\alpha}{1+\alpha^2}\right\} \\
&&\\
M[2]&=&1\\
&&\\
M[3]&=&Ac (3-A^2)\\
&=& c \left\{\frac{\alpha(3+2\alpha^2)}{(1+\alpha^2)^{3/2}}\right\}\\
&&\\
M[4]&=&3\\
&&\\
M[5]&=&c \left\{(Ac)(15 B^4 + 20 A^2 B^2 + 8 A^4)\right\}\\
&=&(Ac)(8 \alpha^4+20 \alpha^2+15)(1+\alpha^2)^{-2}\\
&&\\
M[6]&=&15\\
&&\\
M[7]&=&c\left\{=Ac(105 B^6 + 210 A^2 B^4+168 A^4 B^2 + 48 A^6)\right\}\\
&=&(Ac)(48 \alpha^6+168 \alpha^4+210 \alpha^2+105)(1+\alpha^2)^{-3}\\
&&\\
M[8]&=&105
\end{eqnarray*}

\newpage

\subsection{Description of the test}

\Ni \textbf{(a) p-value}.\\

\Ni For validating hypothesis that the data $X_1$, $\ldots$, $X_n$, for $n\geq 1$, are generated from $SN(\alpha,0,1)$, we can and do use the asymptotic result of Part (2) of Corollary \ref{col1}. Then the significance level of the test, i.e. the $p$-value, is given by

\begin{equation}
pval= \mathbb{F}(\chi_2^2 > j_n), \label{pval}
\end{equation}

\Bin where $j_n$ is the observed value of

\begin{equation}
J_{n}=n \ \left[\Sigma_{22}(a_{n}-a)^{2}+\Sigma_{11}(b_{n}-b)^{2}-2\Sigma_{12}(a_n-a)(b_n-b) \right]/\det(\Sigma). \label{jnobs}
\end{equation}

\Bin \textbf{(b) Numerical implementation}. \label{implement}\\

\Ni In page \pageref{sknPackagage}, we display the numerical implementation as follows.\\

\Ni In (A) We give a value to $\alpha$, then give vectors of mements of $X_1$ and $X_2$ ($M_1$ and $M_2$ respectively) and next compute the moments of
$X \sim SN(0,1,\alpha)$.\\

\Ni In (B), we compute the centered moments up to four, the kurtosis (a) and skewness statistics (b). Next, we give functions $A$ and $B$ as given in Theorem \ref{theo1}, eq. \eqref{A2} and \eqref{B2}.\\

\Ni In (C), we compute the coefficients of the \textit{GJB} statistics under the assumption of a skew normal law of parameter $\alpha$ as fixed. Here we use Monte-Carlo methods. So we make big samples to get the coefficients for once for a fixed value of $\alpha$.\\

\Ni (D), we are ready to proceed to a simulation studies. We fix sample sizes (the smallest the best) a number of repetitions $B$, compute p-values from eq. \eqref{pval} and report the mean p-value to confirm or not that the test actually recognize the true law.\\

\Ni (E) Once Step (D) is proven successful, we can use that validated test to study alternative hypotheses. For $\alpha>0$, i.e. for a non-symmetrical law, we apply the test already set up at Step (C) to samples generated data from the alternative (AH) and check whether or not it rejects it. Here we take the alternative as: (AH) $\alpha=0$, i.e, that data are standard normal. To confirm rejection, it is better to consider the mean $p$-value after a significant number of repetitions $BT$. In a data-driven study, we may appeal to the boostrap method by re-sampling from the sample and get the mean pvalue.\\

\Bin \textbf{(b) Results of our study}.\\

\Ni Let us compute the two main statistics composing the GJBT: the skewness and kurtosis below

\begin{eqnarray*}
be&=&\frac{c(2c^2-1)\alpha^3}{(1+(1-c^2)\alpha^2)^{3/2}}\\
&=&\frac{\sqrt(2)*(4-\pi)*\delta^3}{(\pi-2*\delta^2)^(3/2)}
\end{eqnarray*}

\Ni and

\begin{eqnarray*}
ae&=&\frac{3+6(1-c^2)\alpha^2+(3-2c^2-3c^4)\alpha^4}{(1+(1-c^2)\alpha^2)^{2}}\\
&=&3+\frac{8*(\pi-3)*\delta^4}{(\pi-2*\delta)^2}
\end{eqnarray*}

\Bin For both statistics, the first lines are results of our owns computations above and the second lines stated in \cite{mohsen} but are given in \cite{azzalini1}.\\

\Ni We clearly see how the involvement of $\alpha$ makes the skewness and the kurtosis deviate from those of a  $N(0,1)$ law which are $(0,3)$. The real interest of this paper is the capability of our GJBT to detect that deviation and reject non normality.\\

\Ni We already explain the methodology of the GJBT and its implementation in page \pageref{implement}. From there, we proceed as follows and report the simulation results.\\

\Bin \textbf{(A) Capacity of detecting skew normal data}.\\

\Ni (1) For different values of $\alpha$, we get strong p-values (usually greater than $60\%$ even for low sizes . Here are some examples in Tab. \ref{tabValid}.\\

\begin{table}
	\centering
		\begin{tabular}{r|r|r|r|r|r|r}
		\hline \hline
		$\alpha=$ & $0.1$ & $0.5$ & $1$ & $1.5$& $6$ & $10$ \\
		\hline
		size=2 & $73.21$ & $73.30$ & $75.5$ & $79.79$& $92.21$ & $92.32$ \\
		\hline
			size=10 & $60.77$ & $65.39$ & $64.34$ & $69.9$& $81.74$ & $82.27$ \\
		\hline\hline
	 		\end{tabular}
			\caption{mean p-values by values of $\alpha$ and sizes}
			\label{tabValid}
\end{table}

\Bin \textbf{(G) Rejecting normality, meaning testing $\alpha=0$}.\\

\Ni (2) The test designed to detect $SN(\alpha,0,1)$ data, for $\alpha\neq 0$, does reject the normality of data even for small values of $\alpha$. However, we need a signification size to reach that rejection. In Tab. \ref{tabRjeject1}, we give an idea about the size needed to get the rejection.\\

\begin{table}
	\centering
		\begin{tabular}{r|r|r|r|r|r|r}
		\hline \hline
		$\alpha=$ & $0.1$  & $0.5$ & $1$    & $1.5$ & $6$   & $10$ \\
		\hline
		size     &     $1,000,000$  & $100,000$    & $3,200$ & $750$ & $130$ & $118$ \\
		\hline\hline
			 		\end{tabular}
			\caption{Needed sizes to reject normality}
			\label{tabRjeject1}
\end{table}

\Bin We draw the following conclusion at this point: It seems from here that the \textit{GJBT} works fines to detect non-symmetry and reject symmetry, within the Skew normal family, for relatively small size, for values of $\alpha$ of the order of $5$. Yet, we have the following values of the kurtosis and skewness for $\alpha \in ]0, 5[$ in Tab. \ref{tabRjeject2}. We see that the kurtosis and the skewness significantly deviate for $\alpha$ around and overs $1.5$. Needing large sizes to reject symmetry is not a good thing for the test. Indeed, in many areas, we cannot afford these sizes. However, we can fix that in the following way.

\begin{table}
	\centering
		\begin{tabular}{r|r|r|r|r|r|r|r}
		\hline \hline
		$\alpha=$ & $0.1$ & $0.5$ & $1$ & $1.5$   & $5$ & $6$   & $10$ \\
		\hline
		kurtosis & $3.00001$  		& $3.0069$    & $3.06$& $3.18$ 			& $3.7$ 	& $3.75$ 		& $3.82$ \\
		\hline
	 skewness  & $0.0002$  		& $0.024$    & $0,14$& $0.30$ 			& $0.85$ 	& $0.89$ 		& $0.96$ \\
		\hline\hline
			 		\end{tabular}
			\caption{Values of kurtosis and Skewness for different values of $\alpha$}
			\label{tabRjeject2}
\end{table}

\Bin \textbf{(F) Final recommendations and best practice $\alpha=0$}.\\

\Ni Our simulations show the following things.\\

\Ni (R1) Given a sample $\mathcal{E}_n$ from $SN(\alpha,0,1)$ of any size $n$, we may duplicate $k$ times. We still have a sample of the same law of size ($nk$), that we denote $\mathcal{E}_{k,n}$. By submitting this sample to the test, we still get p-values bigger that $50\%$.\\

\Ni (R2) Now, given data of size $n$, even small, we can do the same by duplicating that data $k$ times. Now, we may choose it big enough to get very small p-values, even null, and reject the hypotheses, because of (R1). Even by taking very big $k$'s, we may afford minutes of waiting completion of the computations to get the rejection or not. However, for $\alpha=0.01$, our simulations show that, even for a value that makes $nk$ higher that 10 millions, we cannot have the rejection. We can understand that by considering Tab. \ref{tabRjeject2} and see 
that $SN(\alpha,0,1)$ the kurtosis and skewness are almost identical between $SN(\alpha,0,1)$ and $\mathcal{N}(0,1)$ for $0\leq \alpha 0.05\leq $. So, for values less than $0.4$, we can accept normality. For values more than $0.5$, we can and do take $k$ until one million to get rejection.\\ 

\Ni We implement Point (G) on page \pageref{sknDuplicatedTest}. First, we get an estimated value of $\alpha$. If it as low as 
$0.5$, we can accept the symmetry, Otherwise, we consider a duplication of the data size, say $L$,  $k$ time to reach $K=k*L$, $K$ may be taken up to one million and apply the test. Following the value of the estimation of $\alpha$, Tab. \ref{tabRjeject1} can indicate us a value of $K$ to use.\\ \label{skwFixk}

\Ni To be able to apply this rule, we have to apply Algorithm (G) (page \pageref{sknDuplicatedTest}) by activating the case (1) where $X$ is the duplication of the true law $SN(\alpha,0,1)$ and check if the p-value (pvale) is good. If successful, we deactivating Case (1) and apply Case (2) to get rejection of the Normal law.\\

\Ni (R3) From Tab. \ref{tabRjeject2}, we see that for $\alpha$ below $0.5$, values for kurtosis and skewness for the normal law are almost equal. So this is recommended : when we get a $95\%$-confidence interval $[c,d]$ and if $c$ is less that $0.5$, we accept the hypothesis. Otherwise, we apply (R2) above.\\

\Ni There is also another empirical rule. The test of the true model gives very good
p-values for sizes $n\geq 2$. To the contrary, for values from $2$, even if the rejection is not obtained for moderate sizes, the p-values do not exceed $50\%$ when $\alpha$ is not zero. So by choosing randomly subsamples of moderate sizes $n$, even arounf $n=5$, we can reject symmetry if the p-values does not exceed some thresholds, say $50\%$.\\

\section{Conclusion} The \textit{GJBT} is a very powerful tool for detecting non-symmetry in the Skew normal family of probability laws. The packages shared here are ready to be used for practitioners.\\

\Ni \textbf{Acknowledgement}. The authors thank the anonymous referee particularly, the associated editor who supervised the publication process and the whole editorial team. All of them helped us to improve the paper a lot. We praise their professionalism.

\newpage
\Ni {Appendix (M1) Moments of the standard half-normal law}.\label{anmoments} The density of the standard half-normal law $|Z_1|$ with 
$Z_1 \sim \mathcal{N}(0,1)$ is\\

$$
f(x)= c \ \exp(-x^2/2) \ 1_{(x\geq 0)},  \ \ c=\sqrt(2/\pi)
$$

\Bin Then $k$-th moment $L_k$ are given as follows. $L_0=1$. For $k\geq 1$,

\begin{eqnarray}
L_k&=& c \int_{0}^{+\infty} x^k \ \exp(-x^2/2) \ dx= c \int_{0}^{+\infty} x^{k-1} \ d\bigg(-x \ \exp(-x^2/2)\biggr) \ dx\\
&=& c \biggr\{\biggr[ -x^{k-1} \ \exp(-x^2/2)\biggr]_{0}^{+\infty} +  (k-1) \int_{0}^{+\infty} x^{k-1} \ d\bigg(-x \ \exp(-x^2/2)\biggr) \ dx\biggr\}\\
\end{eqnarray}

\Bin Now, for $k=1$, we get

\begin{eqnarray}
L_k &=& c \biggr\{[\exp(-x^2/2)]_{0}^{+\infty} +  0\biggr\}\\
&=&c.
\end{eqnarray}

\Bin For $k>1$, we have

\begin{eqnarray}
L_k&=& (k-1) L_{2k}.
\end{eqnarray}

\Ni Then for $k\geq 0$

\begin{eqnarray}
L_2k&=& (2k-1)(2k-3) ... (1) L_{o}\\
&=& \frac{(2k)!}{2^k k!} = \mathbb{E}\biggr(\mathcal{N}(0,1)^{2k}\biggr)\\
\end{eqnarray}

\Bin and for  $k\geq 1$

\begin{eqnarray}
L_{2k_1}&=& (2k) (2k-2) ... (2) L_1-1)(2k-3) ... (2) L_{1}\\
&=& c 2^k k! \ \ \square
\end{eqnarray}

\newpage
\Ni {\bf Appendix (M2) Moments of the standard Skew law}. \label{snmoments}. We have the representation $Z=AZ_1+BZ_2$. We already know that

$$
M_1=(\mathbb{E}Z_2^j, \ 1\leq j \leq 8)= (0, \ \ 1, \ \  0, \ \  3, \ \  0, \ \  15, \ \  0 ,\ \ 105)
$$

\Bin and

$$
M_2=(\mathbb{E}Z_1^j, \ 1\leq j \leq 8)= (c, \ \ 1, \ \  2c, \ \  3, \ \  8c, \ \  15, \ \  48c ,\ \ 105).
$$

\Bin For for $Z_1=|\mathcal{N}(0,1)|$, a simple integration by parts gives

$$
\mathbb{E}Z_1 = \sqrt(2/\pi) \equiv c
$$ 

\Bin and for $k>1$,

$$
\mathbb{E}Z_1^k = (k-1) \mathbb{E}Z_1^{k-2}
$$ 

\Bin So, we get
$$
M_2=(\mathbb{E}Z_1^j, \ 1\leq j \leq 8)= (c, \ \ 1, \ \  2c, \ \  3, \ \  8c, \ \  15, \ \  48c ,\ \ 105).
$$

\Bin So, the eight first moments are got from

\begin{equation} \label{momexpansion}
M[j]=B^j M_2[j] + \sum_1^j b(h,j) \ a^h \ b^{j-h} \ M_1[h] \ M_2[j-h].
\end{equation}

\newpage
\Ni \textbf{(A) Computing the Moments}.\label{sknPackagage}
\begin{lstlisting}
alpha=1
M <- numeric(8); M1 <-numeric(8); M2 <-numeric(8)
(delta=alpha/sqrt(1+alpha^2))
A=delta
B=sqrt(1-A^2)
c=sqrt(2/pi)
M1[1]=c
M1[2]=1
M1[3]=2*c
M1[4]=3
M1[5]=8*c
M1[6]=15
M1[7]=48*c
M1[8]=105
---------
M2[1]=0
M2[2]=1
M2[3]=0
M2[4]=3
M2[5]=0
M2[6]=15
M2[7]=0
M2[8]=105
(M[1]=A*M1[1]+B*M2[1])
for(j in 2:8){
	yokka=((A^j)*M1[j]) + ((B^j)*M2[j])
	for(h in 1:(j-1)){
		yok=imhBin(h,j)*(A^h)*(B^(j-h))*M1[h]*M2[j-h]
		yokka=yokka+yok
	}
	M[j]=yokka
}
\end{lstlisting}

\Ni\textbf{Remark}. We need the binomial combination number function $imhBin(p,n)$, $0\leq p\leq n$. In R, it is $cbinom(p,n)$. If the reader dos not have a package to compute it, a program of it is given in (\textbf{F}) below.\\
\newpage

\Ni \textbf{(B) Computing the functions $A$ and $B$ in Theorem \ref{theo1}, eq. \eqref{A2} and \eqref{B2}}.
\begin{lstlisting}
## Variance
(sigmaC=M[2]-(M[1]^2))
(sigma=sqrt(sigmaC))
### Recall of the coefficients de GJB
A3 <- function(x) h3(x)-M[1]^3*h0(x)+6*M[1]^2*h1(x)
      -3*M[1]*h2(x)-3*M[2]*h1(x)
A4 <- function(x) h4(x)-4*M[1]*h3(x)+6*M[1]^2*h2(x)
      +(-4*M[3]+12*M[1]*M[2]-12*M[1]^3 )*h1(x)
A2 <- function(x) h2(x)+(M[1]^2)*h0(x)-2*M[1]*h1(x)
(a=muX(4,M)/(muX(2,M)^2))
(b=muX(3,M)/(muX(2,M)^(3/2)))
B2 <- function(x) sigma^(-3)*(A3(x)-((3/2)*sigmaC)*muX(3,M)*A2(x))   
C2 <- function(x) (sigma^(-4))*( A4(x) 
     - (2*(sigma^(-2))*muX(4,M)*A2(x)) )
\end{lstlisting}

\Bin \textbf{Remark}. this packages needs the functions $muX(p, M)$ which give the centred moments form the cumulants $M$, using binomial expansions. Whoever needs is directed to see Point (\textbf{G}) below\\

\Ni \textbf{(C) Computing the coefficient of the GJB statistic under the assumption}.
\begin{lstlisting}
#Monte-Carlo
B=10000
N=1000
j=1
varC=0
varB=0
covarCB=0
Z<- numeric()
for(j in 1:B){
	Z=delta*abs(rnorm(N,0,1))+sqrt(1-delta^2)*rnorm(N,0,1)
	varC=varC + (var(C2(Z))/B)
	varB=varB + (var(B2(Z))/B)
	covarCB=covarCB+(cov(C2(Z),B2(Z))/B)
}
varC
varB
covarCB
(detCB=(varC*varB)-(covarCB^2))
\end{lstlisting}

\newpage
\Ni \textbf{(D) Simulation studies}.
\begin{lstlisting}
#test simulation
n=11
pval=0
B=100
for(h in 1:B){
	#generer n observations de la loi courante 
	X=delta*abs(rnorm(n,0,1))+sqrt(1-delta^2)*rnorm(n,0,1)
	(sigmac=var(X))
 	(sigma=sqrt(sigmac))
 	(moyX=mean(X))
	an=sum((X-moyX)^4)/n
	(an=an/(sigma^4))
	 (bn=sum((X-moyX)^3)/n)
	(bn=bn/(sigma^3))
	(jbg=varB*(an-a)^2)
	(jbg=jbg+varC*((bn-b)^2))
	(jbg=jbg-(2*covarCB*(an-a)*(bn-b)))
	(jbgCurrent=n*jbg/detCB)
	pval=pval+((1-pchisq(jbgCurrent,2))/B)
	#pval=pval+(1-pchisq(jbgCurrent,2))
}
pval
\end{lstlisting}

\newpage
\Ni \textbf{(E) Statistical tests (Applied the designed test)}. \label{sknTest}
\begin{lstlisting}
BT=1000
pvale=0
n=100
for(s in 1:BT){
X = rnorm(n,0,1)
(sigmac=var(X))
(sigma=sqrt(sigmac))
(moyX=mean(X))
an=sum((X-moyX)^4)/n
(an=an/(sigma^4))
(bn=sum((X-moyX)^3)/n)
(bn=bn/(sigma^3))
(jbg=varB*(an-a)^2)
(jbg=jbg+varC*((bn-b)^2))
(jbg=jbg-(2*covarCB*(an-a)*(bn-b)))
(jbgCurrent=n*jbg/detCB)
pvale=pvale+(1-pchisq(jbgCurrent,2))/BT
}
pvale
\end{lstlisting}

\newpage
\Ni \textbf{(F) Final test with sample duplications}. \label{sknDuplicatedTest}
\begin{lstlisting}
#L : size of data
# k: number of repetitions of the samples. 
# How to fix k: See second paragaph of (R2) ...
# ... on page \pageref{skwFixk}
# K=L*k
BT=1000
pvale=0
n=100
for(s in 1:BT){
#(1)# X=rep(delta*abs(rn#(1)#orm(K,0,1))
        +sqrt(1-delta^2)*rnorm(K,0,1))
#(2)# X+rep(rnorm(L,0,1),k)
(sigmac=var(X))
(sigma=sqrt(sigmac))
(moyX=mean(X))
an=sum((X-moyX)^4)/K
(an=an/(sigma^4))
(bn=sum((X-moyX)^3)/K)
(bn=bn/(sigma^3))
(jbg=varB*(an-a)^2)
(jbg=jbg+varC*((bn-b)^2))
(jbg=jbg-(2*covarCB*(an-a)*(bn-b)))
(jbgCurrent=K*jbg/detCB)
pvale=pvale+(1-pchisq(jbgCurrent,2))/BT
}
pvale
\end{lstlisting}

\newpage
\Ni \textbf{(G) A re-program of the combination number [cbinion(pp,qq) in R]}.
\begin{lstlisting}
#binomial
binomp=1; pp=1; nn=1
imhBin <- function(pp,nn){
	binomp=(1/factorial(pp))*factorial(nn)*(1/factorial(nn-pp))
	return (binomp)
}
\end{lstlisting}

\Bin \textbf{(H) Function muX(p,M)}.
\begin{lstlisting}
smu=0
q=1
N <-numeric(0)

muX <- function(q,N){
	smu=(-M[1])^q
	for(j in 1:q){
		smu=smu+(cbinom(q, j)*N[j]*(-N[1])^(q-j))
	}
	return (smu)
}
\end{lstlisting}

\label{fin-art}
\end{document}